\documentclass[11pt,onecolumn,preprintnumbers,amsmath,prc,amssymb]{revtex4}

\usepackage{bm}
\usepackage{amsmath}
\usepackage{slashed}

\usepackage{graphicx}
\usepackage{dcolumn}
\usepackage{bm}
\usepackage{xcolor}
\usepackage{pgfgantt}

\usepackage{graphicx}
\usepackage{float}

\begin{document}

\title{Constraints on the fermionic dark matter from observations of neutron stars}

\author{V. Sagun}
\affiliation{CFisUC, Department of Physics, University of Coimbra, Rua Larga P-3004-516, Coimbra, Portugal}
\email{violetta.sagun@uc.pt, eg@student.uc.pt}

\author{E. Giangrandi}
\affiliation{CFisUC, Department of Physics, University of Coimbra, Rua Larga P-3004-516, Coimbra, Portugal}
\email{eg@student.uc.pt}

\author{O. Ivanytskyi}
\affiliation{Institute of Theoretical Physics, University of Wroclaw, 50-204 Wroclaw, Poland}
\email{oleksii.ivanytskyi@uwr.edu.pl}

\author{I. Lopes}
\affiliation{Centro de Astrof\'{\i}sica e Gravita\c c\~ao  - CENTRA, Departamento de F\'{\i}sica, Instituto Superior T\'ecnico - IST, Universidade de Lisboa - UL, Av. Rovisco Pais 1, 1049-001 Lisboa, Portugal}
\email{ilidio.lopes@tecnico.ulisboa.pt}

\author{K. A. Bugaev}
\affiliation{Bogolyubov Institute for Theoretical Physics, Metrologichna str. 14-B, 03143 Kyiv, Ukraine}
\affiliation{Department of Physics, Taras Shevchenko National University of Kyiv, 03022 Kyiv, Ukraine}
\email{bugaev@th.physik.uni-frankfurt.de}

\maketitle
\section{Abstract} 
We study an impact of asymmetric fermionic dark matter on neutron star properties, including tidal deformability, mass, radius, etc. We present the conditions at which dark matter particles tend to form a compact structure in a core of the star or create an extended halo around it. We show that compact core of dark matter leads to a decrease of the total gravitational mass and tidal deformability compared to a pure baryonic star, while presence of a dark matter halo increases those observable quantities. By imposing an existing astrophysical and gravitational wave constraints set by LIGO/Virgo Collaboration together with the recent results on the spatial distribution of dark matter in the Milky Way we determine a new upper limit on the mass and fraction of dark matter particles inside compact stars. Furthermore, we show that the formation of an extended halo around a NS is incompatible with the GW170817 tidal deformability constraint.




\section{Introduction} 
Despite many decades of searches for dark matter (DM) candidates, its nature still remains unclear. In the last years, several experiments have been trying to put tighter constraints on the DM particle properties (e.g. mass and cross section with nuclei), including the recoil experiments that aim to measure a direct evidence of the interaction between the new particle and nuclear targets.
Astrophysics provides additional possibility to test an effect of DM on the macroscopic objects, e.g. stars. Thus, neutron stars (NSs) are one of the most interesting objects in this context, as they may contain a sizeable amount of DM that was accreated throughout their lifetime and/or was present in a protostar cloud. As it was shown, for example in Refs. \cite{2018PhRvD..97l3007E, 2018arXiv180303266N, 2020PhRvD.102f3028, arXiv:2109.03801}, depending on mass, fraction and self-coupling constant of DM particles the NS properties can be drastically altered. As a general scenario, light DM particles tend to form an extended halo around a NS, while heavy particles are more likely create a dense DM core inside a baryon star. Consequently, the outermost radius of the DM-admixed NS will depend on the DM distribution: in the case of a halo formation the outermost radius will exceed the radius of the baryon component, while a core formation will leave the outermost radius to be the baryon one. In addition, not only the radius of the compact star becomes affected by the presence of DM but its mass as well, leading towards lower values in the latter case. Using astrophysical observations of the most massive pulsars, PSR J0348+0432 of mass $2.01\pm 0.04$ M$_{\odot}$ \cite{2013Sci...340..448A} and PSR J0740+6620 of $2.14^{+0.10}_{-0.09}$ M$_{\odot}$  \cite{2019arXiv190406759C}, as a constraint on the maximum mass of NSs we can study DM-admixed compact stars and their ability to reach this limit.

Moreover, in the past few years, gravitational wave (GW) astronomy has allowed us to input additional observational constraints in such models. In particular, by measuring tidal deformability of NSs during their coalescence. This effect can be quantified via the parameter $\Lambda$, which gives us an information about the properties of matter above the normal nuclear density, i.e. Equation of State (EoS) \cite{Abbott1, Abbott2, 2020ApJ...892L...3A}. Combining the LIGO/Virgo results for GW170817 event that give the upper bound on tidal deformability for 1.4 M$_{\odot}$ star $\Lambda_{1.4} \leq 800$ \cite{Abbott1} together with the astrophysical measurements of two heaviest pulsars it is possible to constrain the properties of DM particles, which is the primary goal of this study.

\vspace*{-0.1cm}
\section{Equations of state of dark and baryon matter} 

DM component is modeled as a relativistic Fermi gas of non-interacting particles with the spin $\frac{1}{2}$ and mass $m_{\chi}$. In our study $m_{\chi}$ is a free parameter that we vary in the interval between 100 MeV and 200 GeV. Relying on the estimation performed in Ref. \cite{2018arXiv180303266N} with good approximation a self-interaction between DM particles  could be neglected. 

For the description of baryon matter (BM) we considered a realistic EoS formulated within
the Induced Surface Tension (IST) concept. The IST EoS represents a unified approach that on the equal footing successfully reproduces the nuclear matter ground state properties \cite{2014NuPhA.924...24S}, hadron matter properties, i.e. multiplicities of hadrons measured in heavy-ion collisions \cite{2018NuPhA.970..133B} and the proton flow constraint \cite{2018PhRvC..97f4905I}, as well as matter inside compact stars \cite{2019ApJ...871..157S}. To be able to describe NSs the model was supplemented by the mean field long-range attraction between baryons, an asymmetry between neutrons and protons, conditions of electric neutrality and $\beta$-equilibrium. In the present study we consider the set B proposed in Ref. \cite{2020PhRvD.101f3025S}. This parameterisation gives 
the values of the nuclear asymmetry energy and its slope at saturation density $J = 30.0$ MeV and $L = 93.19$ MeV, respectively. As our study is focused on the high mass NSs the crust is modeled in a simplified way by the polytropic EoS with adiabatic index $\gamma =\frac{4}{3}$. 

In order to study compact stars composed of mixture between BM and DM we consider the two-fluid Tolman-Oppenheimer-Volkof (TOV) equations
formalism, whereas each of the components is modeled as a perfect fluid. Due to the negligibly weak interaction between two components, only interaction through gravity is worth to consider \cite{2018PhRvD..97l3007E}. Accordingly, the energy-momentum tensors of each component are conserved separately that allows their splitting in the TOV equations (an explicit derivation can be find in the Appendix of Ref. \cite{2020PhRvD.102f3028}).
An amount of DM admixed with BM is described by its fraction $f_{\chi}=\frac{M_D(R_{D})}{M_T}$.
Here $M_{T}=M_{B} (R_{B})+M_{D}(R_{D})$ is the total gravitational mass that equals to the sum of masses of BM and DM components, while $R_{B}$ and $R_{D}$ stand for their corresponding radii. 
\vspace*{-0.1cm}
\section{Results} 

\begin{figure}[ht!]
\centering
\vspace*{-0.5cm}
\includegraphics[width=7.0cm]{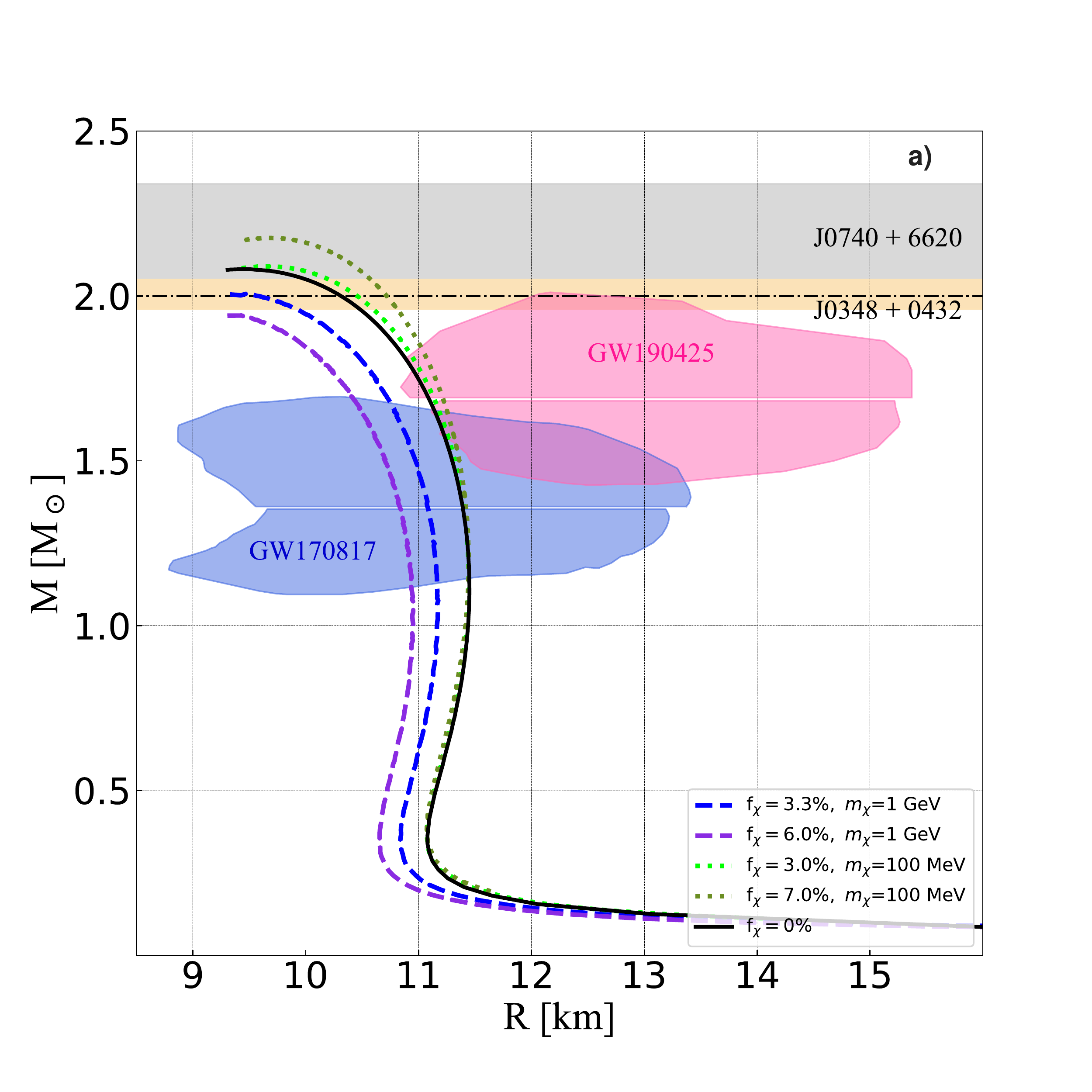}
\includegraphics[width=7.0cm]{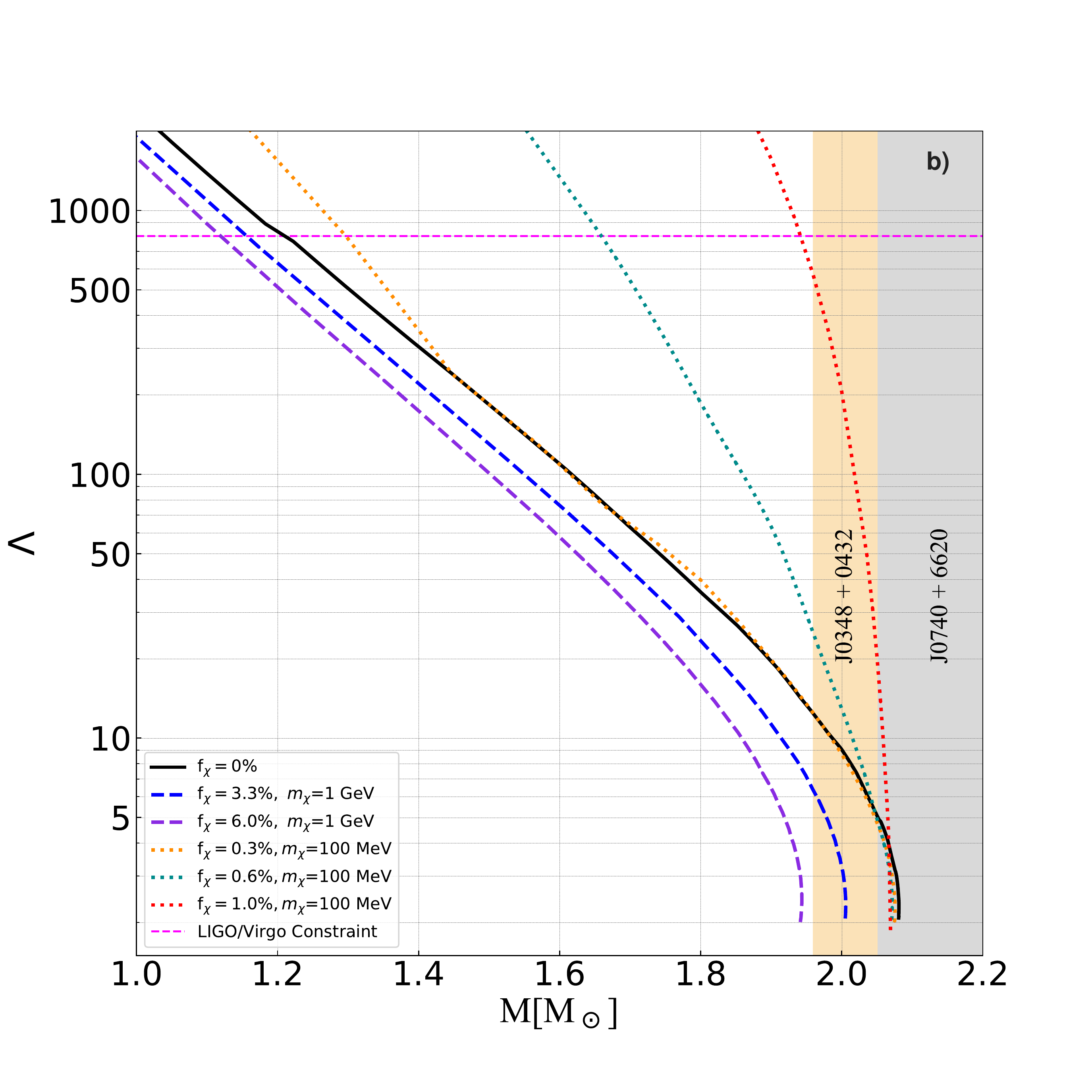} \vspace*{-0.6cm}
\includegraphics[width=7.0cm]{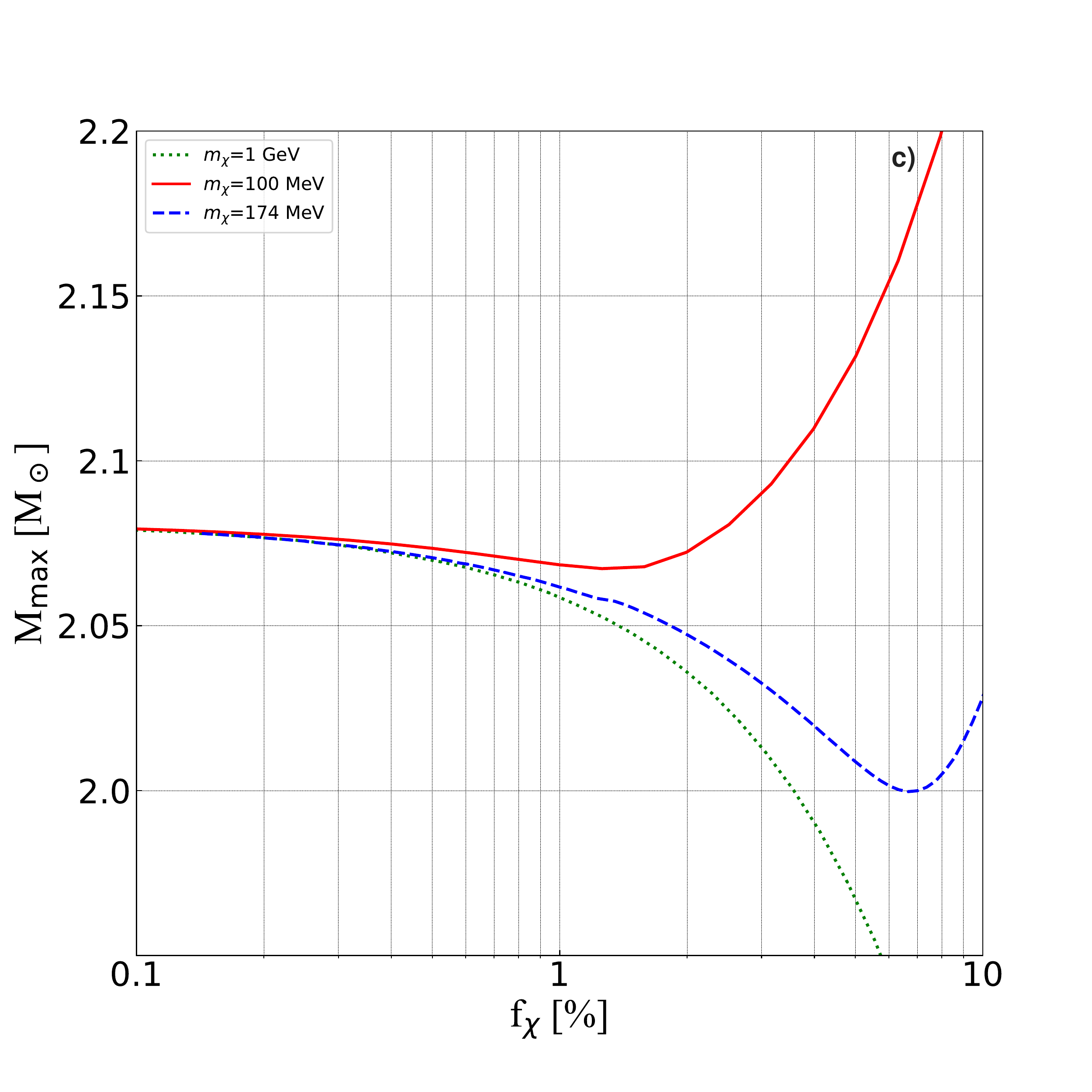}
\includegraphics[width=7.0cm]{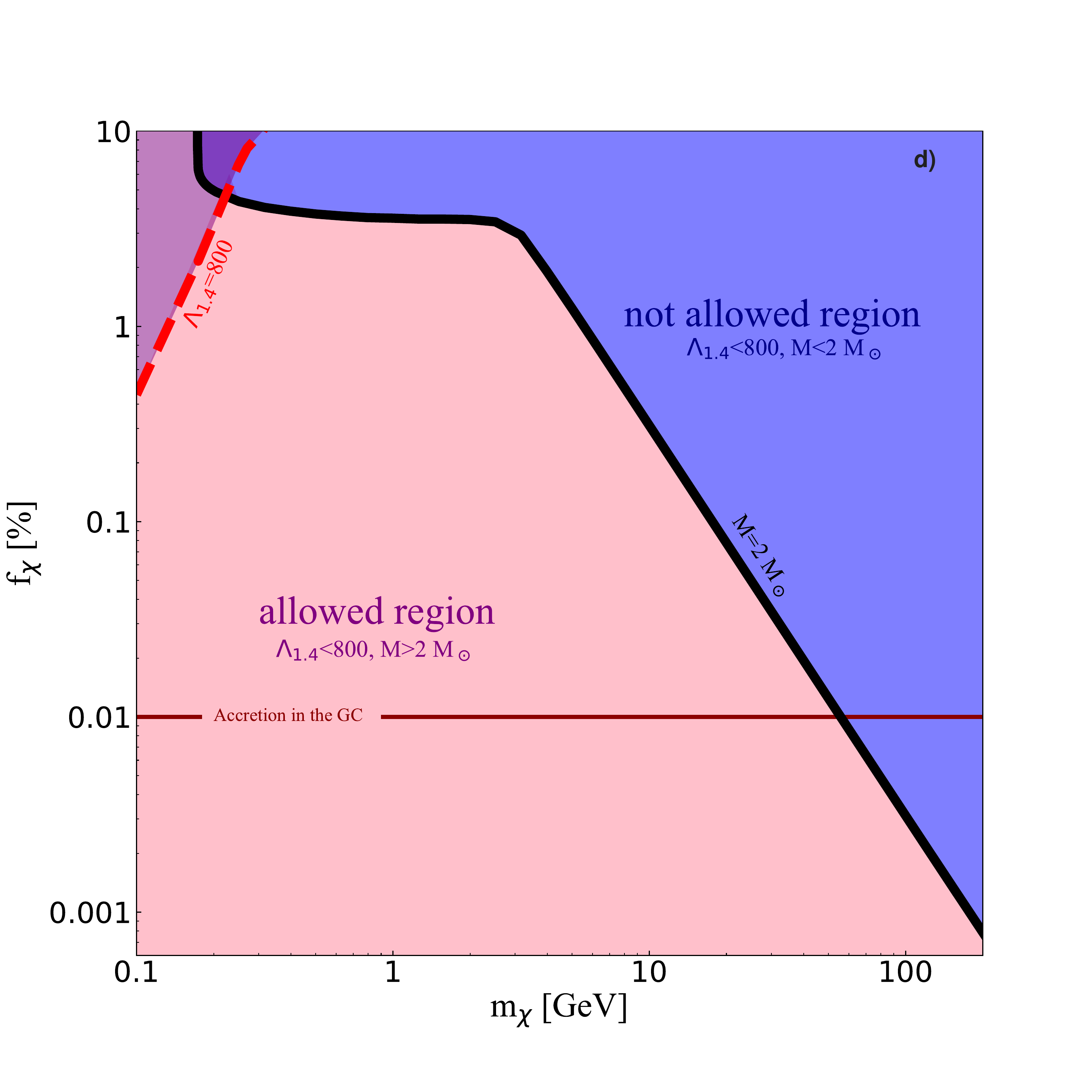}
\caption{Panel (a): Total gravitational mass - baryon radius relations of DM admixed NSs calculated for different masses and fractions of DM. Blue and pink areas represent constraints from GW170817 \cite{Abbott1} and GW190425 \cite{2020ApJ...892L...3A}, respectively. Black dashed horizontal line shows $M_{max} = 2$ M$_{\odot}$ value.  Panel (b): Tidal deformability as a function of total gravitational mass. Magenta dotted line denotes $\Lambda_{1.4}= 800$ constraint \cite{Abbott1}. Black solid curve on panels a) and b) corresponds to pure baryonic stars (without DM). Gray and yellow bands depict two heaviest pulsars \cite{2013Sci...340..448A, 2019arXiv190406759C}. 
Panel (c): Maximal NS mass vs. DM fraction. Panel (d): Fraction of DM as a function of its particle mass. The black and red curves represent the maximum total gravitational mass equal to $M_{max}=2$ M$_\odot$ and tidal deformability  $\Lambda_{1.4}= 800$, respectively. The pink region below the black curve and on the right side from the red curve corresponds to the allowed range of parameters which is in a full agreement with the heaviest known NSs and LIGO/Virgo constraints. An estimation of the fraction of accreated DM into NSs in the central Galaxy region is depicted as the brown line. Its intersection with the black curve gives an upper constraint on the mass of DM particles of 60 GeV (for details see Ref. \cite{2020PhRvD.102f3028}).
}
\label{fig1}
\end{figure}

The mass-radius relations obtained as a solution of the TOV equations are presented on the panel a) in Fig. \ref{fig1}. The black solid curve corresponds to the pure BM IST EoS with maximum NS mass $M_{max} = 2.08$ M$_{\odot}$ and radius $R_{1.4} = 11.37$ km. The dashed blue and violet curves depict the DM-admixed NSs calculated for DM particles with mass $m_{\chi}=1$ GeV and fractions $3.3\%$ and $6.0\%$. The significant compactness of DM-admixed stars in comparison to the BM curve is related to formation of the dense DM core. An opposite case, whereas DM is distributed in a halo is shown with dotted red and olive curves. Light DM particles with mass $m_{\chi}=100$ MeV tend to form an extended halo, e.g. $R_{D}=9.4$ km for $f_{\chi}=0.3\%$
and $R_{D}=135.2$ km for $f_{\chi}=3.0\%$.

\vspace*{-0.3cm}

Depending on whether DM is distributed as a compact core or an extended halo it affects stellar tidal deformibility in a opposite way. We estimate this deformability by the dimensionless parameter $\Lambda=\frac{2}{3} k_2 \left(\frac{R}{M_{T}}\right)^5$, where $R=max(R_{B},R_{D})$ is the outermost radius of DM-admixed NS and $k_2$ is defined according to  Ref. \cite{Hinderer:2007mb}. The tidal deformability calculated for different masses and fractions of DM particles is shown on the panel b). As it is seen, light DM particles with $m_\chi$=0.1 GeV lead to disagreement with the LIGO/Virgo constraint already at fraction $f_{\chi} \gtrsim 0.5\%$.

The panel c) shows the behaviour of $M_{max}$ as a function of $f_\chi$ for DM particles of different mass. In the case of light DM particles this dependence is non monotonous and exhibits a minimum. Note, that growth of $M_{max}$ at high $f_\chi$ is related to formation of the DM halo. For $m_\chi$=0.174 GeV the lowest value of $M_{max}$ equals to exactly $2$ M$_\odot$ that makes this case consistent with the upper mass constraint for any $f_\chi$. For heavier particles, e.g. $m_\chi$=1 GeV (see green dashed curve) the maximum mass is above  $2$ M$_\odot$  only for a certain range of DM fractions.

The panel d) in Fig. \ref{fig1} presents the combined analysis of DM parameters  using the heaviest mass and GW constraints. Thus, black curve represents the values of DM mass and fraction for which $M_{max}=2$ M$_\odot$. This curve splits the range of parameters for the unphysical region coloured in blue where $M_{max}<2$ M$_\odot$ and the pink region with $M_{max}>2$ M$_\odot$. In addition, a purple region on the left side from the red dashed curve is a non-allowed region since it leads to $\Lambda_{1.4}>800$. As a result, the pink area below the black solid and red dashed curves correspond to the allowed region of parameters for which $M_{max}>2$ M$_\odot$ and  $\Lambda_{1.4}<800$.


\vspace*{-0.1cm}
\section{Conclusions} 

We analyze an impact of fermionic DM accumulated in a NS on a stellar radius, mass and tidal deformability. This effect is determined by the total DM mass, or fraction, contained within the baryonic NS, and mass of DM particles. In the presence of a DM core we see more compact star with lower maximum mass and tidal deformability, in comparison to the pure BM NS. However, when the radius of the DM component is larger than the visible NS radius, the tidal deformability value increases significantly, as it depends on the radius of the outermost component, DM halo in this context. In particular, we see that in the presence of an extended DM halo tidal polarisability could no longer be satisfied by the upper bound on $\Lambda_{1.4}$ set by LIGO/Virgo Collaboration. 

Based on it, a combined analysis of the maximum mass of DM-admixed NSs and their tidal deformability led to the formulation of the constraint on the mass and fraction of fermionic DM particles. The $\Lambda_{1.4} \leq 800$ result from GW170817 event rejected a possibly of having an extended halo of light DM particles around compact stars, while the $2$ M$_{\odot}$ limit shaped a range of allowed fractions of DM for higher particle's mass. Furthermore, we demonstrate how our estimate of the fraction of DM accumulated within a NS in the most central part of the Galaxy
combined with future measurements of the $2M_{\odot}$ NS in that region can set the upper constraint on the mass of fermionic DM, which should not exceed 60 GeV.

\vspace*{-0.1cm}
\acknowledgments
V.S., E.G. and I.L. acknowledge the support from the Funda\c c\~ao para a Ci\^encia e Tecnologia (FCT) within the projects UID/04564/2021, UIDB/04564/2020, UIDP/04564/2020 and PTDC/FIS-AST/28920/2017. V.S. also acknowledges the PHAROS COST Action CA16214. The work of O.I. was supported by the Polish National Science Center under the grant No. 2019/33/BST/03059. K.A.B. acknowledges the support from the National Academy of Sciences of Ukraine (Grant number CC-10-2021).

\end{document}